\documentclass[11pt]{article}

\usepackage{latexsym,amsfonts,amsmath,amssymb,wasysym,enumerate,epsfig}
\usepackage{theorem}
\usepackage{mathrsfs}
\usepackage{tikz}
\usepackage{ulem}
\usepackage{bm}
\usepackage[all]{xy}

\usepackage{color}

\theorembodyfont{\rmfamily}

\newcommand{\be}{\begin{equation}}
\newcommand{\ee}{\end{equation}}
\newcommand{\beu}{\begin{equation*}}
\newcommand{\eeu}{\end{equation*}}
\newcommand{\bea}{\begin{eqnarray}}
\newcommand{\eea}{\end{eqnarray}}
\newcommand{\beaa}{\begin{eqnarray*}}
\newcommand{\eeaa}{\end{eqnarray*}}
\newcommand{\bmx}{\begin{pmatrix}}
\newcommand{\emx}{\end{pmatrix}}

\newcommand{\proof}{{\noindent\textbf{Proof. }}}
\newcommand{\finproof}{{\hfill \rule{5pt}{5pt}}}

\newcommand{\CC}{{\mathbb C}}

\def\and{\quad\mbox{and}\quad}

\newcommand{\llangle}{\langle\!\langle}
\newcommand{\rrangle}{\rangle\!\rangle}
\newcommand{\steady}{|{\cal S}\rangle} 
\advance\textheight by \topskip
\begin{document}
\setcounter{page}{0}
\pagestyle{empty}
%
%
\begin{center}

 {\LARGE  {\sffamily  Exact solution to integrable open multi-species SSEP\\[1.2ex]
 and macroscopic fluctuation theory} }\\[1cm]

\vspace{10mm}
  
{\Large 
 M. Vanicat$^{a}$\footnote{matthieu.vanicat@lapth.cnrs.fr}}\\[.41cm] 
{\large $^{a}$  LAPTh, CNRS - Universit{\'e} Savoie Mont Blanc.\\[.242cm]
   9 chemin de Bellevue, BP 110, F-74941  Annecy-le-Vieux Cedex, 
France. }
\end{center}
\vfill

\begin{abstract}
We introduce a multi-species generalization of the symmetric simple exclusion process with open boundaries. This model possesses the property
of being integrable and appears as physically relevant because the boundary conditions can be interpreted as the interaction with particles
reservoirs with fixed densities of each species. The system is driven out-of-equilibrium by these reservoirs.
The steady state is analytically computed in a matrix product form. This algebraic structure allows us to obtain exact expressions for the mean
particle currents and for the one and two-point correlation functions. An additivity principle is also derived from the matrix ansatz and
permits the computation of the large deviation functional of the density profile. We also propose a description of the model in the context of 
the macroscopic fluctuation theory and we check the consistency with the exact computations from the finite size lattice. 
\end{abstract}

\vfill\vfill
\rightline{LAPTh-060/16}
\rightline{October 2016}

\newpage
\pagestyle{plain}

\section{Introduction.}
One important challenge of statistical physics is to understand the non-equilibrium stationary states. Such systems, in which 
the detailed balance is broken, are not governed by a Boltzmann statistics. So far there is no general framework to describe the stationary distribution
at the microscopic level. In this perspective, it appears essential to look for exactly solvable stochastic models and to compute analytically 
their non-equilibrium stationary distribution, in order to put some light on their structure. Finding physically relevant integrable
models is a hard task and often requires dealing with the Yang-Baxter equation and with representation theory of quantum groups.
The symmetric simple exclusion process (SSEP) is a stochastic model which belongs to this privileged category. 
It aims to describe a very simple situation where two particle reservoirs at different densities are connected by a pipe. 
Particles of the highest density reservoir will pour into the lowest density reservoir and the system will display a non vanishing 
particle current in the stationary regime.

More precisely particles diffuse on a one dimensional lattice connected with the two reservoirs at its extremities. A particle can hop 
to the left or right neighboring site with equal rate. An hard-core constraint is imposed: there is at most one particle per site. 
Particles are injected and extracted with asymmetric rates on the boundaries to mimic the coupling with the reservoirs.
It turns out that this model is integrable: the Markov matrix governing the stochastic dynamics is identical to the Hamiltonian of the 
Heisenberg XXX spin chain with open boundaries.
The complete spectrum of the Markov matrix has been computed using algebraic Bethe ansatz \cite{MRM,BC}
and the stationary distribution has been expressed in a matrix product form \cite{DEHP,DerrReview}. 
The latter has revealed to be a very powerful technique to compute observables 
in the stationary state and has permitted a rigorous microscopic verification of the Macroscopic Fluctuation Theory
(MFT) \cite{DLS1,DLS2,DCairns,DerrReview}.

There are at least two ways to extend this model. The first is to add an external field in the bulk which induces an asymmetry in the
hopping rate of the particles. It leads to the Asymmetric Simple Exclusion Process (ASEP) \cite{MartinRev,CMZ}
which also shares the integrability property
and displays a rich physical phenomenology with boundary induced phase transitions \cite{Krug,DEHP,SD}.
The second is to try to extend the model to different
species of particles. This is the goal of the present paper. The main difficulty is to tune properly the particles injection and extraction rates
in order to maintain the integrability of the model. This requires the solution of the reflection equation, which can be seen as the boundary equivalent of 
the Yang-Baxter equation. It turns out that at least one solution of this equation has a nice physical interpretation. The exchange rates at the 
boundaries can be interpreted as an interaction with a particles reservoir which imposes a fixed density for each species.

Let us mention here that the integrable multi-species exclusion processes have received a lot of interest in the last few years. In particular
the steady state of the periodic multi-species ASEP was exactly constructed in a matrix product form \cite{AAMP1,AAMP2,EFM,MMR,PEM} and 
its algebraic structure
intensively studied \cite{CdGW,KMO}. The open boundaries case revealed to be more intricate, but some specific example were studied with
reflexive boundaries \cite{Ari}, semi permeable boundaries \cite{ALS,Uch} or inhomogeneous hopping rate in the bulk \cite{Kar}. A large class of
integrable boundaries was given in \cite{CFRV} and two-species examples were studied in \cite{CEMRV}.
The model we propose to study here is simpler than the aforementioned ones because of the symmetric dynamics in the bulk.
Nevertheless it displays a non-equilibrium stationary state and seems to be of physical interest.
The simple matrix product structure of the steady state allows us to push the computations of the physical observables up to the 
derivation of the large deviation functional of the density profile and allows us to make contact with the MFT.

The plan of the paper reads as follows. In section \ref{sec:presentation}, we present the stochastic dynamics of the model and the associated 
Markov matrix. Then we construct and prove in section \ref{sec:Matrix_Ansatz} a matrix product expression for the stationary state. 
The particular case of the thermodynamic equilibrium is also discussed. In section \ref{sec:observables}, using this algebraic 
framework, we compute physical quantities such as the mean particles currents and the one and two points correlation functions. 
Thanks to the matrix ansatz, we formulate in section \ref{sec:additivity} an additivity principle. It then permits to compute the large 
deviation functional of the density profile. We propose in section \ref{sec:MFT} a description of our model in the framework of MFT. 
We check the consistency with the exact computations from the finite size lattice.
Finally in section \ref{sec:integrability} we recall the integrability framework used 
to deal with open systems and we give without any proof the expression of the reflection matrices associated to our model.

\section{Presentation of the model.\label{sec:presentation}}

\subsection{Dynamical rules.}
We consider a system involving $N$ species of particles 
which diffuse on a one dimensional lattice comprising $L$ sites. Each site can be in $N$ different states $s=1,\dots,N$ 
depending on its occupancy. More precisely, we set $\tau_i=s$
if the site at position $i$ carries a particle of species $s$, with $1\leq s \leq N$.
Hence a configuration of the lattice will be denoted by a $L$-uplet $(\tau_1,\dots, \tau_L) \in \left\{1,\dots,N\right\}^L$.
Let us remark that one of the particle species, for instance species labeled by $1$, may be interpreted as holes on the lattice.

The dynamics is stochastic. During an infinitesimal time $dt$, in the bulk, there is a probability $dt$ that two particles
of different species, located on two adjacent sites, exchange their positions. 
At the left boundary, a particle of species $s'$ located on the first site can be replaced by a particle of species $s$ with 
probability $dt \times \alpha_s/a$. In the same way, at the right boundary,  a particle of species $s'$ located on the last site
can be replaced by a particle of species $s$ with probability $dt \times \beta_s/b$.

 Later on, the parameters $\alpha_1,\dots,\alpha_N$ (respectively $\beta_1,\dots,\beta_N$) will be interpreted as the particle densities 
 at the left (respectively right) reservoirs. We have thus the constraints
 \begin{equation} \label{eq:sum_densities}
  \sum_{s=1}^N \alpha_s =1, \quad \mbox{and} \quad \sum_{s=1}^N \beta_s =1.
 \end{equation} 
 The number $a$ (respectively $b$) will be seen as the distance between the left reservoir 
 and the first site (respectively the distance between the right reservoir and the last site), the lattice spacing being one in the bulk.  

The update rules of the stochastic process described above are summarized in the following table where the rates
of the allowed transitions are depicted above the arrows:
 \begin{equation}
 \begin{array}{|c |c| c| }
 \hline \text{Left} & \text{Bulk} & \text{Right} \\
 \hline
\rule{0pt}{4ex} s'\, \xrightarrow{\ \alpha_s/a\ }\, s&  s's\, \xrightarrow{\ 1\ } \,ss'&s'\, \xrightarrow{\ \beta_s/b\ } \,s\\ [1ex]
 1\leq s,s'\leq N&1\leq s,s' \leq N&1\leq s,s' \leq N\\ \hline
 \end{array}
 \end{equation}
Let us stress that the injection and extraction rate of each species at the boundaries are not the most general. The particular model
presented here is motivated by the fact that it is integrable (see section \ref{sec:integrability}). It turns out that it has a nice physical interpretation.
Taking into account the constraints \eqref{eq:sum_densities}, we are left with $2\times N$ free parameters. 
For a generic choice of these parameters, the system will be driven out of equilibrium by the two reservoirs.

\null

 \noindent \textbf{Remark.} The system will reach, in the long time limit, a thermodynamic equilibrium if and only if the reservoir densities of 
 each species of particle are the same on the left and on the right, namely: $\alpha_s=\beta_s$, for all $1\leq s \leq N$.  The detailed
 balance condition is indeed satisfied only in this case.

\null

\subsection{Markov matrix and master equation.}
In this subsection we set up the mathematical formalism needed to write the probability density function of the model and its time
evolution (master equation) in a concise vector form. This will be also of great help to compute and express in a simple form the 
stationary probability density function.

Let us first attach to each site of the lattice a vector space $\mathbb{C}^N$ with basis $|1\rangle,|2\rangle,\dots,|N\rangle$, where
$|s\rangle=(\underbrace{0,\dots,0}_{s-1},1,\underbrace{0,\dots,0}_{N-s})^t$. The set of all configurations of the lattice is thus 
embedded in $\underbrace{\CC^N\otimes\cdots\otimes\CC^N}_{L}$ with natural basis $|\tau_1\rangle\otimes \cdots \otimes |\tau_L\rangle$,
where $\tau_i=1,2,...,N$.
We denote by $P_t(\tau_1,\dots,\tau_L)$ the probability for the system to be in configuration $(\tau_1,\dots,\tau_L)$ at time $t$. 
These probabilities can be encompassed in a single vector
\begin{equation}
 |P_t\rangle = \begin{pmatrix}
                P_t(1,\dots,1,1) \\
                P_t(1,\dots,1,2) \\
                \vdots \\
                P_t(N,\dots,N,N)
               \end{pmatrix}
             = \sum_{1\leq \tau_1,\dots,\tau_L \leq N} P_t(\tau_1,\dots,\tau_L)  \, |\tau_1\rangle\otimes \cdots \otimes |\tau_L\rangle.
\end{equation}
This allows us to write in a compact form the master equation, governing the time evolution of the probability density
\begin{equation} \label{eq:master_equation}
 \frac{d |P_t\rangle}{dt}=M|P_t\rangle,
\end{equation}
where the Markov matrix $M$ is given by
\begin{equation} \label{eq:Markov_matrix}
 M=B_1+\sum_{i=1}^{L-1}m_{i,i+1} +\overline{B}_L.
\end{equation}
The matrices $B$, $\overline{B}$ and $m$ are the local jump operators. The indices denote the sites,
or equivalently the copies of $\mathbb{C}^N$, on which
the operators act non trivially (they act as the identity in the other copies). The matrix $B$ encodes the dynamics at the left boundary 
and acts on the first site as 
\begin{equation}
 B|s'\rangle =-\frac{1}{a}|s'\rangle+ \sum_{1\leq s\leq N} \frac{\alpha_s}{a}|s\rangle, \qquad 1\leq s'\leq N,
\end{equation}
which leads to the explicit expression 
\begin{equation} \label{eq:mat_B}
 B=\frac{1}{a}\begin{pmatrix}
     \alpha_1-1 & \alpha_1 & \alpha_1 & \hdots & \hdots & \alpha_1 \\
     \alpha_2 & \alpha_2-1 & \alpha_2 & \hdots & \hdots & \alpha_2 \\
     \alpha_3 & \alpha_3 & \alpha_3-1 & \hdots & \hdots & \alpha_3 \\
     \vdots &  \vdots &   & \ddots &  & \vdots \\
     \alpha_{N-1} &  \alpha_{N-1} & \hdots & \hdots &  \alpha_{N-1} -1 &  \alpha_{N-1} \\
     \alpha_N  & \alpha_N & \hdots & \hdots &  \alpha_N & \alpha_N-1
   \end{pmatrix}.
\end{equation}
In the same way, the matrix $\overline B$ encodes the dynamics at the right boundary 
and acts on the last site as 
\begin{equation}
 \overline B|s'\rangle = -\frac{1}{b}|s'\rangle+\sum_{1\leq s\leq N} \frac{\beta_s}{b}|s\rangle,  \qquad 1\leq s'\leq N,
\end{equation}
which leads to the explicit expression 
\begin{equation} \label{eq:mat_Bb}
 \overline B=\frac{1}{b}\begin{pmatrix}
     \beta_1-1 & \beta_1 & \beta_1 & \hdots & \hdots & \beta_1 \\
     \beta_2 & \beta_2-1 & \beta_2 & \hdots & \hdots & \beta_2 \\
     \beta_3 & \beta_3 & \beta_3-1 & \hdots & \hdots & \beta_3 \\
     \vdots &  \vdots &   & \ddots &  & \vdots \\
     \beta_{N-1} &  \beta_{N-1} & \hdots & \hdots &  \beta_{N-1} -1 &  \beta_{N-1} \\
     \beta_N  & \beta_N & \hdots & \hdots &  \beta_N & \beta_N-1
   \end{pmatrix}.
\end{equation}
Finally the matrix $m$ acts on two adjacent sites and encodes the dynamics in the bulk as
\begin{equation} \label{eq:bulk_jump_operator}
 m |s'\rangle \otimes |s\rangle = |s\rangle \otimes |s'\rangle - |s'\rangle \otimes |s\rangle.
\end{equation}
It can be expressed as $m=P-1$, where $P$ is the permutation operator, namely $P|v\rangle \otimes |w\rangle = |w\rangle \otimes |v\rangle$ if 
$|v\rangle,|w\rangle \in \mathbb{C}^N$.

\null

\noindent \textbf{Remark.} The well-known SSEP model with one species of particles plus holes is recovered from this framework for $N=2$ 
 (one has then to identify species $1$ with holes). The present parameters are in this case related to the usual one $\alpha$, $\beta$, 
 $\gamma$ and $\delta$ by
 $\alpha_1=\gamma/(\alpha+\gamma)$, $\alpha_2=\alpha/(\alpha+\gamma)$, $\beta_1=\beta/(\beta+\delta)$, $\beta_2=\delta/(\beta+\delta)$,
 $a=1/(\alpha+\gamma)$ and $b=1/(\beta+\delta)$. Note that this corresponds to the change of variable already used to study the one species SSEP,
 see for instance \cite{DerrReview}.

\null

\section{Matrix product solution.\label{sec:Matrix_Ansatz}}

This section  is devoted to the construction of the stationary state of the model. More precisely we want to compute the vector $\steady$
which satisfies the stationary version of the master equation \eqref{eq:master_equation}, that is $M\steady =0$.
The entries of this vector can be expressed in a matrix product form, that is the probability to observe a configuration 
$(\tau_1,\dots,\tau_L)$ in the steady state can be written as
\begin{equation} \label{eq:matrix_product}
 \mathcal{S}(\tau_1,\dots,\tau_L)=\frac{1}{Z_L}\llangle W|X_{\tau_1}X_{\tau_2}\dots X_{\tau_L} |V\rrangle, 
\end{equation}
where $Z_L=\llangle W|C^L|V\rrangle$ is a normalisation, so that the entries of $\steady$ sum to $1$. We have used the notation
\begin{equation} \label{eq:def_C}
C=X_1+\dots+X_N.
\end{equation}
The matrix ansatz was first introduced in the context of exactly solvable out-of-equilibrium models in \cite{DEHP} to construct exactly 
the steady state of the TASEP. Since then, it has proven to be a very efficient method to solve other models and has been widely used in the literature.
The reader can refer to \cite{MartinRev} for a review. The connection between the matrix product construction of the stationary state and 
the integrability of the related model has been pointed out and explained in \cite{CRV,Sasamoto2}.

\subsection{Algebraic relations.}
For the matrix product state \eqref{eq:matrix_product} to compute the stationary distribution correctly, the operators $X_1,\dots,X_N$
and the boundary vectors $\llangle W|$ and $|V\rrangle$ have to satisfy precise algebraic relations.
The root and the meaning of these relations will be exposed in the subsection \ref{subsec:proof}.
The operators $X_1,\dots,X_N$ belong to a Lie algebra
\footnote{The Lie algebra \eqref{eq:lie_algebra} is not semi-simple since there is an abelian ideal of rank $N-1$ generated by the elements 
$\lambda_1 X_s-\lambda_s X_1$ for $2\leq s \leq N$. Hence it does not belongs to the well known classification of semi-simple Lie algebras.
It could be interesting to study the decomposition into solvable and semi-simple parts of this algebra but this is beyond the scope of this paper.}.
They satisfy the commutation relations
\begin{equation} \label{eq:lie_algebra}
 [X_s,X_{s'}]=\lambda_s X_{s'}-\lambda_{s'} X_s, \qquad 1\leq s,s'\leq N,
\end{equation}
where
\begin{equation}
 \lambda_s=\alpha_s-\beta_s, \qquad 1\leq s\leq N.
\end{equation}
Note that the structure constants $\lambda_s$ of the Lie algebra \eqref{eq:lie_algebra} can be absorbed after a rescaling of the 
generators $X_s \longrightarrow \lambda_s X_s$. However, we will not perform this rescaling in the following because it does not
simplify the computations of physical quantities.

The action of the operators $X_s$ on the left boundary vector $\llangle W|$ is given by
\begin{equation}\label{eq:rel_W}
 \llangle W| \big( \alpha_s C-X_s \big) = a\lambda_s \llangle W|, \qquad 1\leq s\leq N,
\end{equation}
where $C$ is defined in \eqref{eq:def_C}.
Note that these $N$ relations are not all independent (the sum of these equations is trivial),
only $N-1$ are necessary.
In the same way the action of the operators $X_s$ on the right boundary vector $|V\rrangle$ read
\begin{equation} \label{eq:rel_V}
 \big( \beta_s C-X_s \big)|V \rrangle = -b\lambda_s |V \rrangle, \qquad 1\leq s\leq N.
\end{equation}
Again, only $N-1$ of these equations are independent.

Unfortunately, we were not able to find an explicit representation for the operators $X_s$ and the boundary vectors $\llangle W|$ and $|V\rrangle$.
However, we will show that the commutation relations \eqref{eq:lie_algebra} and the relations on the boundary vectors
\eqref{eq:rel_W} and \eqref{eq:rel_V} allow us to compute the currents and correlation functions, see section \ref{sec:observables},
and to prove an additivity principle, see section \ref{sec:additivity}.

\null

\noindent \textbf{Remark.} Once again the matrix ansatz solution of the usual SSEP with one species of particles and holes can be obtained
 for $N=2$, by doing the same change of parameters as mentioned in the remark at the end of section \ref{sec:presentation}, and setting
 $D= X_2/\lambda_2$ and $E=-X_1/\lambda_1=X_1/\lambda_2$. They satisfy $DE-ED=D+E$ and $\llangle W|(\alpha E-\gamma D)=\llangle W|$,
 $(\delta E-\beta D)|V\rrangle = -|V\rrangle$. 

\null

\subsection{Proof of the matrix product form.} \label{subsec:proof}
We now show that the algebraic relations presented above \eqref{eq:lie_algebra}, \eqref{eq:rel_W} and \eqref{eq:rel_V} imply that
the matrix product expression \eqref{eq:matrix_product} gives the stationary state of the model. We need first to define two key vectors
\begin{equation}
 \mathbf X = \begin{pmatrix}
    X_1 \\ \vdots \\ X_N
   \end{pmatrix} \quad \mbox{and} \quad 
\overline{ \mathbf X} = \begin{pmatrix}
    \lambda_1 \\ \vdots \\ \lambda_N
   \end{pmatrix}.  
\end{equation}
They are the building blocks of the algebraic relations presented previously.
They allow us to rewrite the vector $\steady$ in the concise form
\begin{equation} \label{eq:steady}
 \steady=\frac{1}{Z_L}\llangle W| \mathbf X \otimes \mathbf X \otimes \dots \otimes \mathbf X |V\rrangle.
\end{equation}
Then the commutation relations between the $X_s$ \eqref{eq:lie_algebra} can be expressed equivalently as the telescopic relation
\begin{equation} \label{eq:tel}
 m \mathbf X \otimes \mathbf X
= \mathbf X \otimes \overline{ \mathbf X}-\overline{ \mathbf X} \otimes \mathbf X.
\end{equation}
In the same way, the equations on the left boundary \eqref{eq:rel_W} are equivalent to
\begin{equation} \label{eq:tel_W}
 \llangle W| B \mathbf X =  \llangle W| \overline{ \mathbf X},
\end{equation}
and the equations on the right boundary \eqref{eq:rel_V} are equivalent to
\begin{equation} \label{eq:tel_V}
 \overline B \mathbf X |V \rrangle = -\overline{ \mathbf X}|V \rrangle.
\end{equation}
It is known, see for instance \cite{CRV} for a proof, that relations \eqref{eq:tel}, \eqref{eq:tel_W} and \eqref{eq:tel_V} are sufficient to
ensure that the vector \eqref{eq:steady} is the stationary state of the Markov matrix \eqref{eq:Markov_matrix}. Acting with the 
Markov matrix $M$ on $\steady$ leads indeed to a telescopic sum.

\subsection{The thermodynamic equilibrium case.}

We already mentioned that the system reaches a thermodynamic equilibrium if and only if $\alpha_s=\beta_s$ for all $1\leq s\leq N$.
In this case we have $\lambda_s=0$ for all $1\leq s\leq N$, which implies that the operators $X_s$ commute one with each other and can be
chosen proportional to the identity operator. We hence set $X_s:=r_s$, with $r_1,\dots,r_N$ real numbers. It is straightforward to check
that $r_s=\alpha_s=\beta_s$ satisfy the boundary relations \eqref{eq:rel_W} and \eqref{eq:rel_V}. 

The steady state is given by
\begin{equation} \label{eq:steady_thermo}
 \steady=\begin{pmatrix}
          r_1 \\ \vdots \\ r_N
         \end{pmatrix} \otimes 
         \begin{pmatrix}
          r_1 \\ \vdots \\ r_N
         \end{pmatrix} \otimes \dots \otimes 
         \begin{pmatrix}
          r_1 \\ \vdots \\ r_N
         \end{pmatrix}.
\end{equation}
This shows that in the thermodynamic equilibrium, the occupation numbers $\tau_1,\dots,\tau_L$ are independent and 
identically distributed random variables.

\section{Currents and correlation functions.\label{sec:observables}}

The algebraic structure of the stationary state described in section \ref{sec:Matrix_Ansatz} proves very powerful in the computation of physical 
quantities such as the correlation functions and the particle currents. The first step is to evaluate the normalisation $Z_L$.

\subsection{Normalisation}
Assuming that the scalar product of the boundary vectors $\llangle W|V\rrangle =1$, the normalisation of the steady state defined
by $Z_L=\llangle W|C^L|V\rrangle$ is equal to
\begin{equation} \label{eq:normalisation}
 Z_L=\frac{\Gamma(a+b+L)}{\Gamma(a+b)},
\end{equation}
where the gamma function satisfies the functional relation $\Gamma(x+1)=x\Gamma(x)$.
\vspace{0.5cm}

\noindent \textbf{Proof.} 
We first remark that because of constraints \eqref{eq:sum_densities}, we have
\begin{equation}
 \sum_{s=1}^N\lambda_s=\sum_{s=1}^N \alpha_s-\sum_{s=1}^N\beta_s=1-1=0.
\end{equation}
It allows us to compute
\begin{equation}
 [X_s,C]=\sum_{s'=1}^N [X_s,X_{s'}]=\lambda_s \sum_{s'=1}^N X_{s'} -X_s \sum_{s'=1}^N\lambda_{s'},
\end{equation}
and leads to the very useful relation
\begin{equation} \label{eq:com_C}
 [X_s,C]=\lambda_s C, \quad \mbox{or equivalently} \quad X_s C=C(X_s+\lambda_s).
\end{equation}
Using this equality $n$ times we obtain 
\begin{equation} \label{eq:com_Ci}
 X_s C^n=C^n (X_s +n\lambda_s).
\end{equation}
We are now equipped to compute the normalisation 
\begin{eqnarray}
 Z_L & = & \llangle W| C^L |V \rrangle = \frac{a\lambda_1}{\alpha_1}Z_{L-1}+\frac{1}{\alpha_1} \llangle W|X_1 C^{L-1} |V \rrangle \nonumber \\
 & = & \frac{\lambda_1}{\alpha_1}(a+L-1)Z_{L-1}+\frac{1}{\alpha_1} \llangle W|C^{L-1}X_1 |V \rrangle \nonumber \\
 & = & \frac{\lambda_1}{\alpha_1}(a+b+L-1)Z_{L-1}+\frac{\beta_1}{\alpha_1}Z_L. \label{eq:calcul_ZL}
\end{eqnarray}
The first line is obtained thanks to relation \eqref{eq:rel_W} for $s=1$.  
We get the second line through relation \eqref{eq:com_Ci} for $s=1$ and $n=L-1$. 
The last equality is established using \eqref{eq:rel_V} for $s=1$. 
Finally \eqref{eq:calcul_ZL} can be rearranged and leads to the recursive relation
\begin{equation}
 Z_L=(a+b+L-1)Z_{L-1}.
\end{equation}
Keeping in mind that $Z_0=\llangle W|V\rrangle =1$, we can solve the previous relation and we obtain \eqref{eq:normalisation}.
\finproof 

\subsection{Particle currents.}
The mean stationary current of the particles of species $s$ between site $i$ and $i+1$ is defined by
the average algebraic number of particles of species $s$ crossing the bound between sites $i$ and $i+1$ per unit of time:
\begin{equation}
  J_s =  \frac{\llangle W| C^{i-1}X_s(C-X_s)C^{L-i-1} |V \rrangle}{Z_L}-\frac{\llangle W| C^{i-1}(C-X_s)X_sC^{L-i-1} |V \rrangle}{Z_L}.
\end{equation}
Its analytical expression is given by
\begin{equation} \label{eq:current}
 J_s=\frac{\lambda_s}{L-1+a+b},
\end{equation}
which is independent of the site $i$, as expected from the conservation of the particles number in the bulk.
\vspace{0.5cm}

\proof 
\begin{eqnarray*}
 J_s & = &  \frac{\llangle W| C^{i-1}[X_s,C-X_s]C^{L-i-1} |V \rrangle}{Z_L} 
  =   \frac{\llangle W| C^{i-1}[X_s,C]C^{L-i-1} |V \rrangle}{Z_L} 
  =  \lambda_s\frac{Z_{L-1}}{Z_L},
\end{eqnarray*}
where the last equality is obtained thanks to \eqref{eq:com_C}.
Hence using \eqref{eq:normalisation} we get the desired expression \eqref{eq:current}.
\finproof 

\null

\noindent \textbf{Remark.} In the thermodynamic equilibrium case, that is when $\lambda_s=0$ for all $s$, all the particle
 currents vanish, as expected.

\null

\subsection{Correlation functions.} \label{subsec:correlation_functions}
For a given configuration, we set $\rho_s^{(i)}=1$ if there is a particle of species $s$ on the site $i$ and $\rho_s^{(i)}=0$ else. 
The algebraic structure of the steady state, revealed by the matrix product formulation, offers a very efficient framework to 
compute the equal time multi-points correlation functions in the stationary state 
$\langle \rho_{s_1}^{(i_1)}\rho_{s_2}^{(i_2)}\dots \rho_{s_k}^{(i_k)}\rangle $, where $\langle \cdot \rangle$ stands for the expectation 
with respect to the stationary measure. We will compute in this subsection only the one and 
two points correlation functions, which are of particular interest for a physical point of view. In principle closed expressions for 
the higher order correlation functions can also be derived using the computational techniques presented below.

The one point function $\langle \rho_s^{(i)} \rangle$ represents the mean density of particles of a given species $s$ at a given site $i$. 
It can be expressed through the matrix product formalism as
\begin{equation}
 \langle \rho_s^{(i)} \rangle =  \frac{\llangle W| C^{i-1}X_sC^{L-i} |V \rrangle}{Z_L}.
\end{equation}
Using the algebraic structure (see the proof below), it can be reduced to the closed expression
\begin{equation} \label{eq:density}
 \langle \rho_s^{(i)} \rangle =\frac{(b+L-i)\alpha_s+(a+i-1)\beta_s}{a+b+L-1}.
\end{equation}
Note that the density profile is the linear interpolation between the left reservoir with density $\alpha_s$ located at distance $a$ from 
the first site and the right reservoir with density $\beta_s$ located at distance $b$ from the last site. We recover the Fourier law.

The two-point correlation function can also be written in a matrix product form 
\begin{equation} 
 \langle \rho_s^{(i)}\rho_{s'}^{(j)} \rangle=\frac{\llangle W| C^{i-1}X_sC^{j-i-1}X_{s'}C^{L-j} |V \rrangle}{Z_L}.
\end{equation}
It leads to a factorised expression for the connected two-point function
\begin{eqnarray}
\langle \rho_s^{(i)}\rho_{s'}^{(j)} \rangle_c & := & 
\langle \rho_s^{(i)}\rho_{s'}^{(j)} \rangle-\langle \rho_s^{(i)} \rangle \langle \rho_{s'}^{(j)} \rangle \nonumber \\
& = & -\lambda_s\lambda_{s'}\frac{(a+i-1)(b+L-j)}{(a+b+L-1)^2(a+b+L-2)}. \label{eq:2pts_function}
\end{eqnarray}
The formulas \eqref{eq:density} and \eqref{eq:2pts_function} are very similar to the ones derived for the usual one-species SSEP 
\cite{DDR,DerrReview} and appear as direct generalisation for the multi-species case.

\noindent \textbf{Proof of \eqref{eq:density} and \eqref{eq:2pts_function}.}
The mean particle density of species $s$ at site $i$ can be computed using the algebraic structure given by the matrix product form
\begin{eqnarray}
 \langle \rho_s^{(i)} \rangle & = & \frac{\llangle W| C^{i-1}X_sC^{L-i} |V \rrangle}{Z_L} 
  =  (L-i)\lambda_s \frac{Z_{L-1}}{Z_L}+\frac{\llangle W| C^{L-1}X_s |V \rrangle}{Z_L} \\
 & = & (b+L-i)\lambda_s \frac{Z_{L-1}}{Z_L}+\beta_s 
  =  \frac{(b+L-i)\alpha_s+(a+i-1)\beta_s}{a+b+L-1}.
\end{eqnarray}
The second equality is obtained using relation \eqref{eq:com_Ci}. We use then \eqref{eq:rel_V} to get the second line of the equation
and the last equality is established thanks to expression \eqref{eq:normalisation}.

For the two-point function, using again \eqref{eq:com_Ci} and \eqref{eq:rel_V}, we have 
\begin{eqnarray}
 \langle \rho_s^{(i)}\rho_{s'}^{(j)} \rangle & = & \frac{\llangle W| C^{i-1}X_sC^{j-i-1}X_{s'}C^{L-j} |V \rrangle}{Z_L} \\
 & = & \lambda_{s'}(L-j+b) \frac{\llangle W| C^{i-1}X_sC^{L-i-1}|V \rrangle}{Z_L}+\beta_{s'} \langle \rho_s^{(i)} \rangle
 \label{eq:two_points_function_intermediate}
\end{eqnarray}
Replacing $L$ by $L-1$ in the expression \eqref{eq:density} we obtain
\begin{eqnarray}
 \frac{\llangle W| C^{i-1}X_sC^{L-i-1}|V \rrangle}{Z_L} & = & \frac{Z_{L-1}}{Z_L}\frac{(b+L-1-i)\alpha_s+(a+i-1)\beta_s}{a+b+L-2} \\
 & = & \frac{Z_{L-1}}{Z_L}\left(  \langle \rho_s^{(i)} \rangle -\lambda_s\frac{i-1+a}{(L-1+a+b)(L-2+a+b)} \right)
\end{eqnarray}
Substituting back in \eqref{eq:two_points_function_intermediate} leads to
\begin{equation}
 \langle \rho_s^{(i)}\rho_{s'}^{(j)} \rangle = \langle \rho_s^{(i)} \rangle \langle \rho_{s'}^{(j)} \rangle
 -\lambda_s\lambda_{s'}\frac{(a+i-1)(b+L-j)}{(a+b+L-1)^2(a+b+L-2)},
\end{equation}
which concludes the proof.
\finproof

\section{Additivity principle and large deviation of the density profile. \label{sec:additivity}}

\subsection{Additivity principle from matrix ansatz.}

In order to write an additivity principle, we will need some definitions. We define two row vectors of size $N$, encompassing
the particle densities at the two reservoirs
\begin{equation}
 \boldsymbol{\alpha} = (\alpha_1,\dots,\alpha_N) \quad \mbox{and} \quad \boldsymbol{\beta} = (\beta_1,\dots,\beta_N).
\end{equation}
We need also the vector 
\begin{equation}
\overline{\boldsymbol{\rho}}(u)=(1-u)\boldsymbol{\alpha}+u\boldsymbol{\beta}
\end{equation}
which achieves the linear interpolation between $\boldsymbol{\alpha}$ and $\boldsymbol{\beta}$.
 Remark that we have the equality 
 \begin{equation}
 \overline{\rho}_s\left(\frac{i+a-1}{a+b+L-1} \right)= \langle \rho_s^{(i)} \rangle,
 \end{equation}
 where we recall that $\langle \rho_s^{(i)} \rangle$ is the mean particle density of species $s$ at site $i$.
 
We denote by $\mathcal{S}_L(\tau_1,\dots,\tau_L | \boldsymbol{\alpha},\boldsymbol{\beta},a,b)$ the probability of the configuration
$(\tau_1,\dots,\tau_L)$ in the stationary state for a system of size $L$ with boundary parameters
$\boldsymbol{\alpha}$, $\boldsymbol{\beta}$, $a$ and $b$.
Up to now the length of the lattice and the boundary parameters were omitted in the notation $\mathcal{S}(\tau_1,\dots,\tau_L)$
because there was no ambiguity, but this precision will make sense when formulating the additivity principle.
The idea of the additivity principle is to express the stationary weights of a system of size $L$ in terms of the stationary weights of
the two subsystems of size $L'$ and $L''$ (with $L=L'+L''$) obtained by cutting the original system in two parts. 
The results presented in this subsection are heavily inspired by what was done in \cite{DCairns,DerrReview} for the usual SSEP 
(with one species of particles plus holes).  

For the present model, the additivity principle reads
\begin{eqnarray} \label{eq:additivity_principle}
 \mathcal{S}_L(\tau_1,\dots,\tau_L | \boldsymbol{\alpha},\boldsymbol{\beta},a,b) & = & -\frac{\Gamma(a+b+L')\Gamma(L''+1)}{\Gamma(a+b+L)} \
 \oint_{u=1} \frac{du}{2i\pi}\frac{1}{u^{a+b+L'}(1-u)^{L''+1}} \\
 & \times &  \mathcal{S}_{L'}(\tau_1,\dots,\tau_{L'} | \boldsymbol{\alpha},\overline{\boldsymbol{\rho}}(u),a,b) \ 
 \mathcal{S}_{L''}(\tau_{L'+1},\dots,\tau_{L} | \overline{\boldsymbol{\rho}}(u),\boldsymbol{\beta},1-b,b) \nonumber 
\end{eqnarray}
This additivity property can be rewritten using the matrix product formalism. Since the algebraic relations \eqref{eq:lie_algebra},
 \eqref{eq:rel_W} and \eqref{eq:rel_V} involving the operators $X_s$ and 
the boundary vectors $\llangle W|$ and $|V\rrangle$ depend explicitly on the boundary parameters,
we need to introduce some more notations.
 We denote by $\widetilde X_s(u)$, $\llangle \widetilde W(u)|$ and $|\widetilde V(u)\rrangle$ the operators and boundary vectors 
 associated to the system with parameters 
 $\boldsymbol{\alpha}$ and $a$ for the left reservoir and $\overline{\boldsymbol{\rho}}(u)$ and $b$ for the right reservoir.
 Namely they satisfy \eqref{eq:lie_algebra}, \eqref{eq:rel_W} and \eqref{eq:rel_V} where $\beta_s$ has been replaced by
 $(1-u)\alpha_s+u\beta_s$ for all $s$:
 \begin{equation} 
 [\widetilde X_s(u),\widetilde X_{s'}(u)]
 =\widetilde \lambda_s(u) \widetilde X_{s'}(u)-\widetilde \lambda_{s'}(u)\widetilde X_s(u)
 =u\big(\lambda_s \widetilde X_{s'}(u)- \lambda_{s'}\widetilde X_s(u)\big), 
\end{equation}
where
\begin{equation}
 \widetilde \lambda_s(u)=\alpha_s-[(1-u)\alpha_s+u\beta_s]=u\lambda_s, 
\end{equation}
and for the boundaries
\begin{equation}
 \llangle \widetilde W(u)| \big( \alpha_s \widetilde C(u)-\widetilde X_s(u) \big) = au\lambda_s \llangle \widetilde W(u)|,
\end{equation}
and
\begin{equation}
 \big( [(1-u)\alpha_s+u\beta_s] \widetilde C(u)-\widetilde X_s(u) \big) | \widetilde V(u)\rrangle
 = -bu\lambda_s | \widetilde V(u)\rrangle, 
\end{equation}
where
\begin{equation}
 \widetilde C(u)=\widetilde X_1(u)+\dots+\widetilde X_N(u).
\end{equation}

 In the same way we denote by $\widehat X_s(u)$, $\llangle \widehat W(u)|$ and $|\widehat V(u)\rrangle$ the operators and boundary vectors 
 associated to the system with parameters 
 $\overline{\boldsymbol{\rho}}(u)$ and $1-b$ for the left reservoir and $\boldsymbol{\beta}$ and $b$ for the right reservoir.
 Namely they satisfy \eqref{eq:lie_algebra}, \eqref{eq:rel_W} and \eqref{eq:rel_V} where $\alpha_s$ has been replaced by
 $(1-u)\alpha_s+u\beta_s$ for all $s$ and $a$ has been replaced by $1-b$:
  \begin{equation} 
 [\widehat X_s(u),\widehat X_{s'}(u)]
 =\widehat \lambda_s(u) \widehat X_{s'}(u)-\widehat \lambda_{s'}(u)\widehat X_s(u)
 =(1-u)\big(\lambda_s \widehat X_{s'}(u)- \lambda_{s'}\widehat X_s(u)\big), 
\end{equation}
where
\begin{equation}
 \widehat \lambda_s(u)=[(1-u)\alpha_s+u\beta_s]-\beta_s=(1-u)\lambda_s, 
\end{equation}
and for the boundaries
\begin{equation}
 \llangle \widehat W(u)| \big( [(1-u)\alpha_s+u\beta_s] \widehat C(u)-\widehat X_s(u) \big) = (1-b)(1-u)\lambda_s \llangle \widehat W(u)|,
\end{equation}
and
\begin{equation}
 \big( \beta_s \widehat C(u)-\widehat X_s(u) \big) | \widehat V(u)\rrangle
 = -b(1-u)\lambda_s | \widehat V(u)\rrangle, 
\end{equation}
where
\begin{equation}
 \widehat C(u)=\widehat X_1(u)+\dots+\widehat X_N(u).
\end{equation}
 
 We have the formula:
\begin{eqnarray} \label{eq:additivity_principle_MA}
 \llangle W|X_{\tau_1}\dots X_{\tau_L}|V\rrangle & = &  -\oint_{u=1} \frac{du}{2i\pi}\frac{1}{u^{a+b+L'}(1-u)^{L''+1}} \\
 & \times &  \llangle \widetilde W(u)|\widetilde X_{\tau_1}(u)\dots \widetilde X_{\tau_{L'}}(u)|\widetilde V(u)\rrangle \ 
 \llangle \widehat W(u)|\widehat X_{\tau_{L'+1}}(u)\dots \widehat X_{\tau_{L}}(u)|\widehat V(u)\rrangle. \nonumber 
\end{eqnarray}

\noindent \textbf{Proof of \eqref{eq:additivity_principle_MA}.}
For $i=1,\dots,L'$ we perform the change of variables
\begin{equation}
X_{\tau_i}=\alpha_{\tau_i}C-L_{\tau_i} \quad \mbox{and} \quad 
\widetilde X_{\tau_i}(u)=\alpha_{\tau_i}\widetilde C(u)-\widetilde L_{\tau_i}(u).
\end{equation}
The new operators $L_s$ and $\widetilde L_s(u)$ behave conveniently on the left boundary
\begin{equation}
 \llangle W|L_s=a\lambda_s \llangle W| \quad \mbox{and} \quad
\llangle \widetilde W(u)| \widetilde L_s(u)=ua\lambda_s \llangle \widetilde W(u)|.
\end{equation}
When we expand the product $X_{\tau_1}\dots X_{\tau_{L'}}$ (respectively the product $\widetilde X_{\tau_1}(u)\dots \widetilde X_{\tau_{L'}}(u)$),
we can push the $L_s$ (respectively the $\widetilde L_s(u)$) to the left through the $C$'s (respectively the $\widetilde C(u)$'s) using  
the relation $[L_s,C]=-\lambda_s C$ (respectively the relation $[\widetilde L_s(u),\widetilde C(u)]=-u\lambda_s \widetilde C(u)$).
At the end the expansion of $X_{\tau_1}\dots X_{\tau_{L'}}$ involve monomials of the form 
$\lambda_{s_1}\dots \lambda_{s_k}L_{s_{k+1}}\dots L_{s_{n'}}C^{L'-n'}$. The expansion of the product 
$\widetilde X_{\tau_1}(u)\dots \widetilde X_{\tau_{L'}}(u)$ is exactly the same but with the previous monomial replaced by 
$u^k\lambda_{s_1}\dots \lambda_{s_k}\widetilde L_{s_{k+1}}(u)\dots \widetilde L_{s_{n'}}(u)\widetilde C(u)^{L'-n'}$.

In the same way for $i=L'+1,\dots,L'+L''$ we perform the change of variables 
\begin{equation}
X_{\tau_i}=\beta_{\tau_i}C-R_{\tau_i} \quad \mbox{and} \quad  
\widehat X_{\tau_i}(u)=\beta_{\tau_i}\widehat C(u)-\widehat R_{\tau_i}(u).
\end{equation}
The new operators $R_s$ and $\widehat R_s(u)$ behave conveniently on the right boundary
\begin{equation} 
R_s|V\rrangle=-b\lambda_s |V\rrangle \quad \mbox{and} \quad
\widehat R_s(u)|\widehat V(u)\rrangle=-(1-u)b\lambda_s |\widehat V(u)\rrangle.
\end{equation}
Following the same idea as previously, the expansion of $X_{\tau_{L'+1}}\dots X_{\tau_{L}}$ involve monomials of the form 
$\lambda_{s_1}\dots \lambda_{s_k}C^{L''-n''}R_{s_{k+1}}\dots R_{s_{n''}}$. The expansion of the product 
$\widetilde X_{\tau_{L'+1}}(u)\dots \widetilde X_{\tau_{L}}(u)$ is exactly the same but with the previous monomial replaced by \newline 
$(1-u)^k\lambda_{s_1}\dots \lambda_{s_k}\widehat C(u)^{L''-n''}\widehat R_{s_{k+1}}(u)\dots \widehat R_{s_{n''}}(u)$.

Putting all these expansions together, we see that finally it remains to prove
\begin{eqnarray}
 \llangle W|C^{L'+L''-n'-n''}|V\rrangle & = & -\oint_{u=1} \frac{du}{2i\pi}\frac{1}{u^{a+b+L'-n'}(1-u)^{1+L''-n''}} \\
 & \times & \llangle \widetilde W(u)|\widetilde C(u)^{L'-n'}|\widetilde V(u)\rrangle
 \llangle \widehat W(u)|\widehat C(u)^{L''-n''}|\widehat V(u)\rrangle. \nonumber 
\end{eqnarray}
This is established using result \eqref{eq:normalisation} and the fact that
\begin{eqnarray}
 \oint_{u=1} \frac{du}{2i\pi}\frac{1}{u^{a+b+L'-n'}(1-u)^{1+L''-n''}} & = & 
 -\frac{(-1)^{L''-n''}}{(L''-n'')!}\left. \frac{d^{L''-n''}}{du^{L''-n''}} \frac{1}{u^{a+b+L'-n'}} \right|_{u=1} \\
 & = & -\frac{\Gamma(a+b+L'+L''-n'-n'')}{\Gamma(a+b+L'-n')\Gamma(1+L"-n")}.
\end{eqnarray}

\finproof

\subsection{Large deviation of the density profile.}

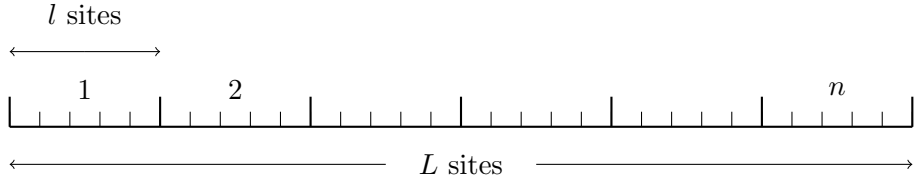
\begin{figure}[htb]
\begin{center}
 \begin{tikzpicture}[scale=1.0]
\draw[thick] (0,0) -- (12,0) ;
\foreach \i in {0,2,...,12}
{\draw[thick] (\i,0) -- (\i,0.4) ;}
\foreach \i in {0,0.4,...,12}
{\draw (\i,0) -- (\i,0.2) ;}
\draw[->] (5,-0.5)--(0,-0.5);
\draw[->] (7,-0.5)--(12,-0.5);
\node at (6,-0.5) [] {$L$ sites};
\draw[->] (1,1)--(0,1);
\draw[->] (1,1)--(2,1);
\node at (1,1.5) [] {$l$ sites};
\node at (1,0.5) [] {$1$};
\node at (3,0.5) [] {$2$};
\node at (11,0.5) [] {$n$};
 \end{tikzpicture}
 \end{center}
 \caption{The system of length $L$ is divided into $n$ boxes of length $l$.}
 \label{fig:subdivision}
\end{figure}

We are interested in evaluating the probability of observing in the stationary state a given density profile in the limit of large system size $L$.
In order to formalize the problem, we split the full system which contains $L=nl$ sites into $n$ subsystems (called ``boxes'' below)
containing $l$ sites each, see figure \ref{fig:subdivision}.

We denote by $P_L(\{\boldsymbol{\rho}^{\{1\}},\boldsymbol{\rho}^{\{2\}},\dots,\boldsymbol{\rho}^{\{n\}}\}\ |\ \boldsymbol{\alpha},\boldsymbol{\beta})$
the probability to find $\rho_s^{\{k\}}\times l$ particles \footnote{the notation $\rho_s^{\{k\}}$ (which stands for the mean number of particles
of species $s$ in the box $k$) should not be confused with the occupation number $\rho_s^{(i)}$ defined in subsection \ref{subsec:correlation_functions}
(which is equal to $1$ if there is a particle of species $s$ at site $i$ and $0$ else). The link between these variables is 
$\rho_s^{\{k\}}=\frac{1}{l}\sum_{i=kl+1}^{(k+1)l}\rho_s^{(i)}$.} 
of species $s$ in the box $k$ for $s=1,\dots,N$ and $k=1,\dots,n$. The row vector
$\boldsymbol{\rho}^{\{k\}}$ encompasses the particles densities of each species in the box $k$:
\begin{equation}
 \boldsymbol{\rho}^{\{k\}}=(\rho_1^{\{k\}},\dots,\rho_N^{\{k\}}).
\end{equation}
Note that we have $\rho_1^{\{k\}}+\dots+\rho_N^{\{k\}}=1$, for all $1\leq k \leq n$.

 For $L$ large we expect the probability 
 $P_L(\{\boldsymbol{\rho}^{\{1\}},\boldsymbol{\rho}^{\{2\}},\dots,\boldsymbol{\rho}^{\{n\}}\}\ |\ \boldsymbol{\alpha},\boldsymbol{\beta}) $ to follow 
 a large deviation principle
\begin{equation}
 P_L(\{\boldsymbol{\rho}^{\{1\}},\boldsymbol{\rho}^{\{2\}},\dots,\boldsymbol{\rho}^{\{n\}}\}\ |\ \boldsymbol{\alpha},\boldsymbol{\beta})\sim
 \exp\left(-L\mathcal{F}_n(\{\boldsymbol{\rho}^{\{1\}},\boldsymbol{\rho}^{\{2\}},\dots,\boldsymbol{\rho}^{\{n\}}\}\ |\ \boldsymbol{\alpha},\boldsymbol{\beta})\right).
\end{equation}
In the limit where the number of boxes $n\gg 1$ and the size of each box $l\gg 1$, we can define a continuous coordinate $x$ such that 
$k=xL$ and a vector $\boldsymbol{\rho}(x)=\boldsymbol{\rho}^{\{k\}}$. We obtain in this case a large deviation functional 
$\mathcal{F}(\{ \boldsymbol{\rho}(x)\}\ |\ \boldsymbol{\alpha},\boldsymbol{\beta})$
\begin{equation}
 P_L(\{\boldsymbol{\rho}(x)\}\ |\ \boldsymbol{\alpha},\boldsymbol{\beta})\sim
 \exp\left(-L\mathcal{F}(\{\boldsymbol{\rho}(x)\}\ |\ \boldsymbol{\alpha},\boldsymbol{\beta})\right).
\end{equation}

In the particular case of the thermodynamic equilibrium, i.e. when $\boldsymbol{\alpha}=\boldsymbol{\beta}:=\boldsymbol{r}=(r_1,\dots,r_N)$, 
see \eqref{eq:steady_thermo}, the large deviation functional is given by
\begin{equation} \label{eq:large_dev_thermo}
 \mathcal{F}(\{ \boldsymbol{\rho}(x)\}\ |\ \boldsymbol{r},\boldsymbol{r})= \int_0^1 dx \ B(\boldsymbol{\rho}(x),\boldsymbol{r}),
\end{equation}
where
\begin{equation}
 B(\boldsymbol{\rho},\boldsymbol{r})= \sum_{s=1}^{N} \rho_s\ln \left(\frac{\rho_s}{r_s}\right)
\end{equation}
We recall that $r_1+\dots+r_N=1$ and $\rho_1(x)+\dots+\rho_N(x)=1$ for all $x$. Remark that $ B(\boldsymbol{\rho}(x),\boldsymbol{r})$ is 
nothing else but the Kullback-Leibler divergence between the two discrete measure $\boldsymbol{\rho}(x)$ and $\boldsymbol{r}$.
\vspace{0.5cm}

\noindent \textbf{Proof of \eqref{eq:large_dev_thermo}.}
In the thermodynamic equilibrium case the stationary distribution is given by \eqref{eq:steady_thermo}. Hence we can 
easily evaluate
\begin{equation}
 P_L(\{\boldsymbol{\rho}^{\{1\}},\boldsymbol{\rho}^{\{2\}},\dots,\boldsymbol{\rho}^{\{n\}}\}\ |\ \boldsymbol{r},\boldsymbol{r})=
 \prod_{k=1}^n \frac{l!}{(l\rho_1^{\{k\}})!\dots (l\rho_N^{\{k\}})!} r_1^{l\rho_1^{\{k\}}}\dots r_N^{l\rho_N^{\{k\}}}.
\end{equation}
Then using the Stirling formula we obtain
\begin{equation}
 \lim\limits_{l\rightarrow \infty} -\frac{1}{L}
\ln P_L(\{\boldsymbol{\rho}^{\{1\}},\boldsymbol{\rho}^{\{2\}},\dots,\boldsymbol{\rho}^{\{n\}}\}\ |\ \boldsymbol{r},\boldsymbol{r})=
 \frac{1}{n}\sum_{k=1}^n \sum_{s=1}^N \rho_s^{\{k\}}\ln \left(\frac{\rho_s^{\{k\}}}{r_s}\right).
\end{equation}
The limit of large $n$ thus gives
\begin{equation}
 \lim\limits_{n\rightarrow \infty} \lim\limits_{l\rightarrow \infty} -\frac{1}{L}
 \ln P_L(\{\boldsymbol{\rho}^{\{1\}},\boldsymbol{\rho}^{\{2\}},\dots,\boldsymbol{\rho}^{\{n\}}\}\ |\ \boldsymbol{r},\boldsymbol{r})=
 \int_0^1 dx \ \sum_{s=1}^{N} \rho_s(x)\ln \left(\frac{\rho_s(x)}{r_s}\right),
\end{equation}
which yields the desired result.
\finproof 
\vspace{0.5cm}

The non-equilibrium case $\boldsymbol{\alpha} \neq \boldsymbol{\beta}$ is more involved: 
\begin{equation} \label{eq:large_dev}
 \mathcal{F}(\{ \boldsymbol{\rho}(x)\}\ |\ \boldsymbol{\alpha},\boldsymbol{\beta})= 
 \int_0^1 dx \ \Big[B(\boldsymbol{\rho}(x),\overline{\boldsymbol{\rho}}(u(x)))+ \ln u'(x)\Big],
\end{equation}
where
$u$ is the monotonic solution of the differential equation 
\begin{equation} \label{eq:diff_eq_F}
 \frac{u''(x)}{(u'(x))^2}+\sum_{s=1}^N \lambda_s \frac{\rho_s(x)}{\overline{\rho}_s(u(x))}=0
\end{equation}
satisfying boundary conditions $u(0)=0$ and $u(1)=1$.

We can deduce from this expression that the most probable density profile is given by $\boldsymbol{\rho}(x)=\overline{\boldsymbol{\rho}}(x)$.
The differential equation is indeed solved by the function $u(x)=x$ in this case because $\lambda_1+\dots+\lambda_N=0$.
Injecting in \eqref{eq:large_dev} makes the large deviation function vanish.

\null

\noindent \textbf{Remark.} The thermodynamic equilibrium case can be of course recovered from the general case. Indeed we have 
 $\overline{\boldsymbol{\rho}}(u)=\boldsymbol{\alpha}=\boldsymbol{\beta}=\boldsymbol{r}$ for all $u$. 
 Moreover the differential equation \eqref{eq:diff_eq_F} reduces to $u''(x)=0$ because $\lambda_s=0$ for all $s$ in this case.
 It is solved by the function $u(x)=x$. Injecting in \eqref{eq:large_dev} leads to \eqref{eq:large_dev_thermo} as expected.

\null

\noindent \textbf{Proof of \eqref{eq:large_dev}.}
The proof presented here follows heavily the lines of the proof written in \cite{DCairns,DerrReview} for the one species SSEP.
For the sake of simplicity, we will present the proof for the case where $a+b=1$, but the generalisation to the other cases is straightforward.

We want to evaluate the probability 
$P_L(\{\boldsymbol{\rho}^{\{1\}},\boldsymbol{\rho}^{\{2\}},\dots,\boldsymbol{\rho}^{\{n\}}\}\ |\ \boldsymbol{\alpha},\boldsymbol{\beta})$
to find $\rho_s^{\{k\}}\times l$ particles of species $s$ in the box $k$ for $s=1,\dots,N$ and $k=1,\dots,n$. This is done by summing 
the probabilities of all the configurations satisfying these constraints. For each of these configurations, we use the additivity 
principle \eqref{eq:additivity_principle} to divide the system into two part of size $L'=kl$ (containing $k$ boxes) and $L''=(n-k)l$
(containing $n-k$ boxes), for a fixed $1\leq k\leq n$. We thus obtain
\begin{equation}
\begin{aligned}
& P_{nl}\left(\{\boldsymbol{\rho}^{\{1\}},\dots,\boldsymbol{\rho}^{\{n\}}\}\, |\, \boldsymbol{\alpha},\boldsymbol{\beta}\right)
  =  -\frac{(kl)!((n-k)l)!}{(nl)!} \
 \oint \frac{du}{2i\pi}\frac{1}{u^{kl+1}(1-u)^{(n-k)l+1}} \\
& \hspace{3cm} \times  P_{kl}\left(\{\boldsymbol{\rho}^{\{1\}},\dots,\boldsymbol{\rho}^{\{k\}}\}\, |\, \boldsymbol{\alpha},\overline{\boldsymbol{\rho}}(u)\right) \, 
P_{(n-k)l}\left(\{\boldsymbol{\rho}^{\{k+1\}},\dots,\boldsymbol{\rho}^{\{n\}}\}\, |\, \overline{\boldsymbol{\rho}}(u),\boldsymbol{\beta}\right) 
\end{aligned}
\end{equation}
In the large $l$ limit, evaluating the previous expression at the saddle point,
we obtain the following equation for the large deviation function 
\begin{equation}
\begin{aligned}
 & \mathcal{F}_n\left(\{\boldsymbol{\rho}^{\{1\}},\dots,\boldsymbol{\rho}^{\{n\}}\}\, |\, \boldsymbol{\alpha},\boldsymbol{\beta}\right)
 = \max\limits_{0<u<1} \quad \frac{k}{n}\ln\left(\frac{nu}{k}\right)+\frac{n-k}{n}\ln\left(\frac{n(1-u)}{n-k}\right)\\
& + \frac{k}{n} \mathcal{F}_k\left(\{\boldsymbol{\rho}^{\{1\}},\dots,\boldsymbol{\rho}^{\{k\}}\}\, |\, \boldsymbol{\alpha},\overline{\boldsymbol{\rho}}(u)\right)
+ \frac{n-k}{n}\mathcal{F}_{n-k}\left(\{\boldsymbol{\rho}^{\{k+1\}},\dots,\boldsymbol{\rho}^{\{n\}}\}\, |\, \overline{\boldsymbol{\rho}}(u),\boldsymbol{\beta}\right)
\end{aligned}
\end{equation}
We repeat $n$ times the same procedure to obtain 
\begin{equation}
 \begin{aligned}
  & \mathcal{F}_n\left(\{\boldsymbol{\rho}^{\{1\}},\dots,\boldsymbol{\rho}^{\{n\}}\}\, |\, \boldsymbol{\alpha},\boldsymbol{\beta}\right)
 = \max\limits_{0=u_0<u_1<\dots<u_n=1} \frac{1}{n} \sum_{k=1}^n 
 \mathcal{F}_1\left(\boldsymbol{\rho}^{\{k\}}\, |\, \overline{\boldsymbol{\rho}}(u_{k-1}),\overline{\boldsymbol{\rho}}(u_k)\right) \\
 & \hspace{10cm} +\ln\left((u_{k}-u_{k-1})n\right)
 \end{aligned}
\end{equation}
In the large $n$ limit, we can define the continuous variable $x=k/n$ and a function $u$ such that $u(x)=u_k$. The sequence $u_k$ being monotone,
the difference $u_k-u_{k-1}$ is small in this limit. Hence we have that $\overline{\boldsymbol{\rho}}(u_{k-1})\simeq \overline{\boldsymbol{\rho}}(u_k)$
and we can replace $\mathcal{F}_1\left(\boldsymbol{\rho}^{\{k\}}\, |\, \overline{\boldsymbol{\rho}}(u_{k-1}),\overline{\boldsymbol{\rho}}(u_k)\right)$
by the equilibrium value 
$\mathcal{F}_1\left(\boldsymbol{\rho}^{\{k\}}\, |\, \overline{\boldsymbol{\rho}}(u_k),\overline{\boldsymbol{\rho}}(u_k)\right)=
B\left(\boldsymbol{\rho}^{\{k\}}\, |\, \overline{\boldsymbol{\rho}}(u_k)\right)$. We thus obtain
\begin{equation}
 \mathcal{F}(\{ \boldsymbol{\rho}(x)\}\ |\ \boldsymbol{\alpha},\boldsymbol{\beta})= 
\max\limits_{u(x)} \int_0^1 dx \ \Big[B(\boldsymbol{\rho}(x),\overline{\boldsymbol{\rho}}(u(x)))+ \ln u'(x)\Big],
\end{equation}
where the maximum is evaluated over the increasing functions $u$ satisfying $u(0)=0$ and $u(1)=1$.
The Euler-Lagrange equation associated with the maximization over $u$ of this functional gives the differential equation \eqref{eq:diff_eq_F}.
\finproof

Let us stress that exact computation, from 
finite size lattice, of the large deviation functional of the density profile has only be achieved on a few out-of-equilibrium models,
including the SSEP \cite{DLS1,DLS2} and the ASEP \cite{DLS3,DLS4}.

\section{Macroscopic fluctuation theory. \label{sec:MFT}}

\subsection{Hydrodynamic description of the multi-species SSEP.}

The macroscopic fluctuation theory (MFT) is a general approach that aims to describe out of equilibrium diffusive particle gases in the 
thermodynamic limit. It was 
developed a few years ago by Bertini, De Sole, Gabrielli, Jona-Lasinio and Landim \cite{Bertini1,Bertini2}, and has proven to be  an efficient way 
to compute fluctuations of the current and of the density profile. One strength of this theory is to describe the diffusive systems through only 
two key parameters, the diffusion constant $D(\rho)$ and the conductivity $\sigma(\rho)$ which depend on the local particle density $\rho$.
These parameters can be determined case by case from the microscopic dynamics of the model.
See \cite{GianniRevue} for a detailed review. Some validations from a 
microscopic point of view were realised for exactly solvable models including the SSEP \cite{DLS1,DLS2,DCairns,DerrReview}, 
and more recently a dissipative system \cite{CRRV}. All these works were related to systems with a single species of particles.

In this section we propose, based on the exact microscopic computations, a hydrodynamic description of the multi-species SSEP
which extends the MFT to systems with several species of particles. We check the consistency with the exact results of the previous sections, 
the rigorous proof of the approach remains to be done (and lies beyond the scope of the present paper).

We define, when $L\rightarrow \infty$, the macroscopic density $\rho_s(x,t)$ of the species $s$ at time $t$ \footnote{Note that
in order to do this hydrodynamic limit, we have rescaled  in all this section the time appearing in \eqref{eq:master_equation} with a factor $L^2$, 
as usual in this context.}
and at position $x\in[0,1]$ 
on the lattice by
\begin{equation}
 \rho_s(x,t)\simeq \frac{1}{2\varepsilon L}\sum_{\lvert i-Lx\rvert\leq L\varepsilon} \rho_s^{(i)},
\end{equation}
where $\varepsilon$ tends to zero and $L\varepsilon$ tends to infinity in the thermodynamic limit.
We denote by $Q_s^{t,(i,i+1)}$ the algebraic number of particles of species $s$
that have crossed the bound between sites $i$ and $i+1$ during the time interval $[0,t]$. It allows us to consider 
\begin{equation}
 Q_s(x,t)\simeq \frac{1}{2\varepsilon L^2}\sum_{\lvert i-Lx\rvert\leq L\varepsilon} Q_s^{t,(i,i+1)}.
\end{equation}
The macroscopic particle current $j_s(x,t)$ of species $s$ at time $t$ and at position $x$ is then defined as 
\begin{equation}
 j_s(x,t)=\frac{\partial}{\partial t}Q_s(x,t).
\end{equation}

\subsection{Rate function for the multi-species SSEP.}

The idea of the MFT is to express the probability to observe certain density profiles $\boldsymbol{\rho}(x,t)=(\rho_1(x,t),\dots,\rho_N(x,t))$
and current profiles $\boldsymbol{j}(x,t)=(j_1(x,t),\dots,j_N(x,t))$ during the time interval $[t_1,t_2]$ as a large deviation principle.
We present now one of the main result of this paper, which gives a new perspective on the rate function of diffusive models with exclusion which
can be seen as that of a model of free particles but with an additional constraint
\begin{equation} \label{eq:proba_MFT}
 P\left(\{\boldsymbol{\rho}(x,t),\boldsymbol{j}(x,t)\}\right) \sim
 \exp \left[ -L \int_{t_1}^{t_2} dt \int_0^1 dx \sum_{s=1}^N \frac{(j_s(x,t)+\partial_x\rho_s(x,t))^2}{4\rho_s(x,t)} \right],
\end{equation}
where the fields satisfy the usual conservation law 
\begin{equation}
 \frac{\partial }{\partial t}\boldsymbol{\rho}(x,t)=-\frac{\partial }{\partial x}\boldsymbol{j}(x,t),
\end{equation}
the boundary conditions
\begin{equation} \label{eq:boundary_conditions}
\boldsymbol{\rho}(0,t)=\boldsymbol{\alpha}, \qquad \boldsymbol{\rho}(1,t)=\boldsymbol{\beta}
\end{equation}
and the additional constraints
\begin{equation} \label{eq:constraints}
\rho_1(x,t)+\dots+\rho_N(x,t)=1, \qquad j_1(x,t)+\dots+j_N(x,t)=0.
\end{equation}
The rate function \eqref{eq:proba_MFT} can be heuristically interpreted having in mind that, for Brownian particles,
the diffusion constant is $D(\rho)=1$ and the conductivity is $\sigma(\rho)=2\rho$. The functional \eqref{eq:proba_MFT} is exactly the one that describes
a model of independent Brownian particles of $N$ different species, but on top of that we impose the exclusion constraint \eqref{eq:constraints} which 
translates the fact that there is at most one particle per site. We recall that in our notation the holes (empty sites) are interpreted as a species of
particles. This formula is supported by: (i) the consistency check with the large deviation
functional of the density profile in the stationary state done in the next subsection, (ii) the following remark. 

\null

\noindent \textbf{Remark.} The well known case of the SSEP with a single species and holes can be recovered from \eqref{eq:proba_MFT} by setting $N=2$.
 Indeed, if we assume that species $1$ plays the role of holes and species $2$ the role of particles,
 we have in this case $j_2(x,t)=-j_1(x,t)=j(x,t)$ and $\rho_2(x,t)=1-\rho_1(x,t)=\rho(x,t)$ due to the constraints \eqref{eq:constraints}.
 Then the rate function in \eqref{eq:proba_MFT} becomes 
\begin{equation} 
\int_{t_1}^{t_2} dt \int_0^1 dx \frac{(j(x,t)+\partial_x\rho(x,t))^2}{4\rho(x,t)(1-\rho(x,t))},
\end{equation}
 which agrees with the known expression for the single species SSEP (recall that the diffusion constant is given by $D(\rho)=1$ 
 and the conductivity by $\sigma(\rho)=2\rho(1-\rho)$).
 
\null

\subsection{Check with the large deviation of the density profile.}

Following what was done in \cite{DCairns,DerrReview}, this framework allows us to express the probability to observe at time $\tau$ a density profile 
$\boldsymbol{\rho}(x)$ in the stationary state. We have to identify how this deviation is produced, i.e. we have to find 
the optimal path $\boldsymbol{\rho}(x,t)$ such that $\boldsymbol{\rho}(x,-\infty)=\overline{\boldsymbol{\rho}}(x)$ and 
$\boldsymbol{\rho}(x,\tau)=\boldsymbol{\rho}(x)$:
\begin{equation} \label{eq:large_devia_MFT}
 \mathcal{F}(\{ \boldsymbol{\rho}(x)\}\, |\, \boldsymbol{\alpha},\boldsymbol{\beta})=
 \min\limits_{\boldsymbol{\rho}(x,t),\boldsymbol{j}(x,t)}
 \int_{-\infty}^{\tau} dt \int_0^1 dx \sum_{s=1}^N \frac{(j_s(x,t)+\partial_x\rho_s(x,t))^2}{4\rho_s(x,t)}.
\end{equation}
Note that the probability to observe a deviation in the density profile $\boldsymbol{\rho}(x)$ does not depend on the time on which this deviation 
occurs. It means that \eqref{eq:large_devia_MFT} does not depend on $\tau$.
\begin{equation} \label{eq:large_dev_diff}
\mathcal{F}(\{ \boldsymbol{\rho}(x)\}\, |\, \boldsymbol{\alpha},\boldsymbol{\beta})=
 \min\limits_{\delta\boldsymbol{\rho}(x),\boldsymbol{j}(x)} \left[
 \mathcal{F}(\{ \boldsymbol{\rho}(x)-\delta\boldsymbol{\rho}(x)\}\, |\, \boldsymbol{\alpha},\boldsymbol{\beta})+
 \delta\tau \int_0^1 dx \sum_{s=1}^N \frac{(j_s(x)+\rho_s'(x))^2}{4\rho_s(x)}\right],
\end{equation}
where we have used the definitions $\boldsymbol{\rho}(x)-\delta\boldsymbol{\rho}(x)=\boldsymbol{\rho}(x,\tau-\delta\tau)$ and
$\boldsymbol{j}(x)=\boldsymbol{j}(x,\tau)$. The conservation law reads $\delta\boldsymbol{\rho}(x)=-\boldsymbol{j}'(x)\times \delta\tau$.
If we define 
\begin{equation}
 U_s(x)=\frac{\delta  \mathcal{F}(\{ \boldsymbol{\rho}(x)\}\, |\, \boldsymbol{\alpha},\boldsymbol{\beta})}{\delta \rho_s(x)},
\end{equation}
we can write using \eqref{eq:large_dev_diff} an equation satisfied by the $U_s(x)$'s. Indeed, maximising \eqref{eq:large_dev_diff} over
the current profile $\boldsymbol{j}(x)$ with the constraint \eqref{eq:constraints} yields
\begin{equation}
 j_s(x)=-\rho_s'(x)+2\rho_s(x)U_s'(x)-2\rho_s(x)\mu(x),
\end{equation}
with the Lagrange multiplier
\begin{equation}
 \mu(x)=\sum_{s=1}^N\rho_s(x)U_s'(x).
\end{equation}
Using the fact that $\sum_{s=1}^{N}j_s(0)U_s(0)=\sum_{s=1}^{N}j_s(1)U_s(1)=0$ (because of the boundary conditions \eqref{eq:boundary_conditions}),
we can perform an integration by part 
and derive an equation satisfied by the functions $U_s'(x)$
\begin{equation} \label{eq:Hamilton}
 \int_0^1 dx \left[\sum_{s=1}^N \left(\rho_s'(x)U_s'(x)-\rho_s(x)U_s'(x)^2 \right)
 +\left(\sum_{s=1}^N\rho_s(x)U_s'(x)\right)^2\right] =0.
\end{equation}
We can check that the large deviation functional exactly computed in section \ref{sec:additivity} indeed fulfills this equation.
We deduce from \eqref{eq:large_dev} that
\begin{equation}
 U_s(x)=\ln\left(\frac{\rho_s(x)}{\overline{\rho}_s(u(x))}\right)+1,
\end{equation}
where the function $u$ satisfies \eqref{eq:diff_eq_F}. Using the constraints \eqref{eq:constraints} and the expression of $U_s(x)$, 
the differential equation \eqref{eq:diff_eq_F} can be rewritten 
\begin{equation}
 \frac{u''(x)}{u'(x)}=-\sum_{s=1}^N\rho_s(x)U_s'(x).
\end{equation}
This permits to show that 
\begin{equation}
 \left(\frac{u''}{u'}\right)'(x)=\sum_{s=1}^N \left(\rho_s'(x)U_s'(x)-\rho_s(x)U_s'(x)^2\right)+\left(\sum_{s=1}^N \rho_s(x)U_s'(x) \right)^2.
\end{equation}
Then we deduce that the left hand side of \eqref{eq:Hamilton} is equal to
\begin{equation}
 \int_0^1 dx\left(\frac{u''}{u'}\right)'(x)= \frac{u''(1)}{u'(1)}-\frac{u''(0)}{u'(0)}=0,
\end{equation}
because $u''(1)=u''(0)=0$ thanks to \eqref{eq:diff_eq_F}.

Let us stress that the Hamilton-Jacobi equation \eqref{eq:Hamilton} obtained when trying to compute the large deviation functional 
of the density profile from the MFT formalism has only been solved in a few models including the weakly asymmetric simple exclusion process 
\cite{ED,DPS} and the Kipnis-Marchioro-Presutti model \cite{KMP,BGL}.

\section{Integrability. \label{sec:integrability}}

\subsection{Integrability in a nutshell.}
In this section we summarize the integrability framework in which the multi-species SSEP with boundaries takes part.
As already mentioned, the Markov matrix \eqref{eq:Markov_matrix} that governs the stochastic dynamics of the model is integrable.
It means that the Markov matrix $M$ belongs to a set of commuting operators which are all encompassed in a generating function $t(x)$ called 
the transfer matrix. The commutation between the operators is ensured by the key property 
\begin{equation} \label{eq:comm_transfer}
[t(x),t(y)]=0.
\end{equation}
For systems with open boundaries \cite{sklyanin}, the transfer matrix is build from two fundamental blocks, the $R$-matrix and the $K$-matrices. 
We present them in the case of the multi-species SSEP. 

The $R$-matrix $\check R(x)=(xP+1)/(x+1)$ acting in $\mathbb{C}^N \otimes \mathbb{C}^N$, where $P$ is the permutation operator defined 
right after \eqref{eq:bulk_jump_operator}, satisfies the braided Yang-Baxter equation 
\begin{equation} \label{eq:ybe}
 \left(\check R(x)\otimes 1 \right) \left( 1 \otimes \check R(x+y) \right) \left(\check R(y) \otimes 1 \right)
 =  \left( 1 \otimes \check R(y) \right) \left(\check R(x+y)\otimes 1 \right) \left( 1 \otimes \check R(x) \right).
\end{equation}
This equation holds in $\mathbb{C}^N \otimes \mathbb{C}^N \otimes \mathbb{C}^N$. The local jump operator $m$ defined in \eqref{eq:bulk_jump_operator}
is intimately linked to $\check R$ through $m=\check R'(0)$.

The $K$ matrices $K(x)=1+2xaB/(x+a)$ for the left boundary and $\overline K(x)=1+2xb\overline B/(x-b)$ for the right boundary,
where $B$ and $\overline B$ are the boundary matrices defined in \eqref{eq:mat_B} and \eqref{eq:mat_Bb}, both satisfy 
the braided reflection equation 
\begin{equation}\label{eq:re}
 \check R(x-y) \left(K(x)\otimes 1\right) \check R(x+y)  \left(K(y)\otimes 1\right) \ = \ 
 \left(K(y)\otimes 1\right) \check R(x+y) \left(K(x)\otimes 1\right) \check R(x-y)
\end{equation}
This equation holds in $\mathbb{C}^N \otimes \mathbb{C}^N$. The boundary matrices $B$ and $\overline B$ are obtained through 
$B=K'(0)/2$ and $\overline B=-\overline K'(0)/2$.

These objects allow us to construct the double row transfer matrix \cite{sklyanin}  
\begin{equation} \label{eq:transfer_matrix}
 t(x)=tr_0\left(\widetilde K_0(x)R_{0,L}(x)\dots R_{0,1}(x)K_0(x)R_{1,0}(x)\dots R_{L,0}(x) \right),
\end{equation}
where $R(x)=P.\check R(x)$ and $\widetilde K(x)=tr_0\left(\overline K_0(-x)((R_{01}(2x)^{t_1})^{-1})^{t_1}P_{01}\right)$. Like the Markov matrix, the transfer matrix acts in
$\underbrace{\CC^N\otimes\cdots\otimes\CC^N}_{L}$. The indices in \eqref{eq:transfer_matrix} indicate the copies of 
$\mathbb{C}^N$ in the tensor space in which the matrices are acting. Remark that we used an additional copy of $\mathbb{C}^N$ with number $0$
which is traced out.

The Markov matrix $M$ is then simply given by $M=t'(1)/2$. The commutation property of the transfer matrix \eqref{eq:comm_transfer} is 
ensured by the Yang-Baxter equation \eqref{eq:ybe} and the reflection equation \eqref{eq:re}. The proof can be found in \cite{sklyanin}.
A more detailed review of the integrable formalism for open exclusion processes can be found in \cite{CRV}.

\subsection{More integrable boundaries.}
As already mentioned, the boundary conditions of the model studied in this paper are specific solutions of the reflection equation \eqref{eq:re}.
These solutions were indeed of particular interest because of their very simple physical interpretation. Nevertheless there exist several other 
solutions to \eqref{eq:re} which provide other integrable stochastic boundary matrices $B$ and $\overline B$.
The solutions of the reflection equation \eqref{eq:re} has been classified in \cite{MRS}. We present here, without proof, classes of stochastic boundaries
among this classification.  
We divide the $N$ species into $p$ distinct families $F_1,\dots,F_p$ of  non-vanishing cardinalities $f_1,\dots,f_p$ at the left boundary 
and into $q$ distinct families $G_1,\dots,G_q$ of non-vanishing cardinalities $g_1,\dots,g_q$
at the right boundary. We hence have two different partitions $\{1,\dots,N\}=\bigsqcup_{k=1}^p F_k=\bigsqcup_{k=1}^q G_k$.
We define $2N$ non negative numbers $\alpha_1,\dots,\alpha_N$ for the left boundary and $\beta_1,\dots,\beta_N$ for the right boundary with
the constraints
\begin{equation}
 \mbox{for all }1\leq k \leq p, \quad  \sum_{s\in F_k} \alpha_s=1,
\end{equation}
and
\begin{equation}
 \mbox{for all }1\leq k \leq q, \quad  \sum_{s\in G_k} \beta_s=1.
\end{equation}

The left boundary conditions are given by
\begin{equation} \label{eq:left_general_boundary}
 B|s'\rangle =-\frac{1}{a}|s'\rangle+ \sum_{s\in F_k} \frac{\alpha_s}{a}|s\rangle,  \qquad 1\leq s' \leq N.
\end{equation}
$k$ in \eqref{eq:left_general_boundary} is such that $s'\in F_k$.
Remark that in the particular case where the family of $s'$ contains only one species, i.e. $F_k=\{s'\}$,
we get from the constraints that $\alpha_{s'}=1$ and hence $ B|s'\rangle = 0$.

In the same way, the right boundary conditions are given by
\begin{equation}
 \overline B|s'\rangle = -\frac{1}{b}|s'\rangle+\sum_{s\in G_k} \frac{\beta_s}{b}|s\rangle,  \qquad 1\leq s' \leq N,
\end{equation}
with $k$ such that $s'\in G_k$.
Note that when we have a single family on the left and a single family on the right, 
i.e. when $p=q=1$, then the boundary conditions reduce to the one studied
in details in this paper. To illustrate these boundary conditions, we give some examples in the case $N=4$ for the left boundary:
\begin{eqnarray}
 B & = & \frac{1}{a}\begin{pmatrix}
  \alpha_1-1 & \alpha_1 & 0 & \alpha_1 \\
  \alpha_2 &\alpha_2-1 & 0 & \alpha_2 \\
  0 & 0 & 0 & 0 \\
  \alpha_4 & \alpha_4 & 0 & \alpha_4-1
 \end{pmatrix}, \quad \mbox{with} \quad \alpha_1+\alpha_2+\alpha_4=1, \label{eq:mat1} \\
 B & = & \frac{1}{a}\begin{pmatrix}
  \alpha_1-1 & \alpha_1 & 0 & 0 \\
  \alpha_2 &\alpha_2-1 & 0 & 0 \\
  0 & 0 & \alpha_3-1 & \alpha_3 \\
  0 & 0 & \alpha_4  & \alpha_4-1
 \end{pmatrix}, \quad \mbox{with} \quad \alpha_1+\alpha_2=1 \mbox{ and } \alpha_3+\alpha_4=1, \\
  B & = & \frac{1}{a}\begin{pmatrix}
  \alpha_1-1 & 0 & \alpha_1 & 0 \\
  0 & \alpha_2-1 & 0 & \alpha_2 \\
  \alpha_3 & 0 & \alpha_3-1 & 0 \\
  0 & \alpha_4 & 0 & \alpha_4-1
 \end{pmatrix}, \quad \mbox{with} \quad \alpha_1+\alpha_3=1 \mbox{ and } \alpha_2+\alpha_4=1, \\
   B & = & \frac{1}{a}\begin{pmatrix}
  0 & 0 & 0 & 0 \\
  0 & \alpha_2-1 & \alpha_2 & 0 \\
  0 & \alpha_3 & \alpha_3-1 & 0 \\
  0 & 0 & 0 & 0
 \end{pmatrix}, \quad \mbox{with} \quad \alpha_2+\alpha_3=1, \label{eq:mat4}  
\end{eqnarray}
Examples of right boundaries in the case $N=4$ are obtained by replacing $\alpha_i$ by $\beta_i$ and $a$ by $b$ in the matrices 
\eqref{eq:mat1}-\eqref{eq:mat4} above.

\section{Conclusion}

In this work we have introduced a multi-species generalization of the SSEP with open boundaries. The boundary conditions are carefully tuned to maintain the 
integrability of the model and are physically interpreted as the coupling with particles reservoirs with fixed densities of each species.
We have expressed analytically the steady state using the matrix ansatz technique and it turns out that the matrices involved belong to a 
simple Lie algebra. Using this convenient algebraic structure we derived exact expressions of physical quantities including the large 
deviation functional of the density profile. We proposed a description of the model in the framework of the MFT and check the consistency 
with the exact computations done on the finite size lattice.

There is of course a lot of work that remains to be done on this model. One thing that is of prime interest in the context of out-of-equilibrium 
statistical physics is to study the fluctuations of the particles currents. The usual way to do that is to perform a current counting deformation
of the Markov matrix. It would be of interest to extend to this multi-species model the results on the current fluctuations obtained 
for the one species open SSEP \cite{DDR,DerrReview}. We could also directly compute the fluctuations of the current from the MFT formalism,
assuming that the density profile which produces an atypical current is time independent. In particular we could try to see if the 
additivity formalism developed in \cite{BD} can be extended to a model with several species of particles. 
The results should be compared with the one given by the current counting deformation of the Markov matrix.

It could be also interesting to study in detail other combinations of integrable boundaries given in section \ref{sec:integrability}. In particular,
following the general approach of the matrix ansatz for integrable systems developed in \cite{Sasamoto2} and \cite{CRV}, it should be possible 
to construct the non-equilibrium stationary state in a matrix product form. It may be enlightening 
to see how the matrix product solution is modified by the change of boundary conditions and whether the additivity principle holds. A description in 
the context of the MFT of such models also remains to be done.

\section*{Acknowledgments}
We are grateful to Caley Finn for helpful discussions. We would like to warmly thank Nicolas Cramp{\'e}, Martin Evans, Kirone Mallick and Eric Ragoucy
for their interest and suggestions.

\end{document}